\documentclass[aps,10pt,prb,twocolumn,letterpaper,superscriptaddress,floatfix]{revtex4-2}

\usepackage{graphicx} 
\usepackage{dcolumn} 
\usepackage{bm} 
\usepackage{siunitx} 
\usepackage{physics} 
\usepackage{xcolor}
\definecolor{PRLblue}{rgb}{0.18,0.18,0.57}
\usepackage[colorlinks=true,allcolors=blue]{hyperref} 

\usepackage{fancyhdr}
\usepackage[version=4]{mhchem}

\begin{document}


\title{Nanosecond dynamics in intrinsic topological insulator \ce{Bi_{2-x}Sb_xSe3} revealed by time-resolved optical reflectivity}


\author{Adam L. Gross}
\affiliation{Department of Physics and Astronomy, University of California, Davis, CA 95616, USA}
\author{Yasen Hou}%
\affiliation{Department of Physics and Astronomy, University of California, Davis, CA 95616, USA}
\author{Antonio Rossi}%
\affiliation{Department of Physics and Astronomy, University of California, Davis, CA 95616, USA}
\affiliation{Advanced Light Source, Lawrence Berkeley National Lab, Berkeley, 94720, USA}
\author{Dong Yu}%
\affiliation{Department of Physics and Astronomy, University of California, Davis, CA 95616, USA}
\author{Inna M. Vishik}%
\email[]{ivishik@ucdavis.edu}
\affiliation{Department of Physics and Astronomy, University of California, Davis, CA 95616, USA}

\date{\today}

\begin{abstract}

\ce{Bi2Se3} is an ideal three-dimensional topological insulator in which the chemical potential can be brought into the bulk band gap with antimony doping. Here, we utilize ultrafast time-resolved transient reflectivity to characterize the photoexcited carrier decay in \ce{Bi_{2-x}Sb_xSe3} nanoplatelets. We report a substantial slowing of the bulk carrier relaxation time in bulk-insulating \ce{Bi_{2-x}Sb_xSe3} as compared to $n$-type bulk-metallic \ce{Bi2Se3} at low temperatures, which approaches $\SI{3.3}{\nano\second}$ in the zero pump fluence limit. This long-lived decay is correlated across different fluences and antimony concentrations, revealing unique decay dynamics not present in $n$-type \ce{Bi2Se3}, namely the slow bimolecular recombination of bulk carriers.

\end{abstract}

\maketitle

Three-dimensional topological insulators (3D TIs) have a nominally insulating bulk and metallic, Dirac-like surface states with spin-momentum-locking \cite{Zhang2009,Moore2010_birth,Hasan2010,Ando2013}. The unique electronic structure of these materials lends itself to numerous electronic, spintronic, or optoelectronic applications \cite{Seradjeh2009,Garate2010,Kong2011,Peng2012,Pesin2012,Kastl2015,Jiang2016,Yue2017,Wang2017,Hou2019,Hou2020}. While the surface states are often the target of these applications, the properties of the bulk are also crucial, particularly for phenomena involving optical excitations which are predominantly initiated in the bulk. Due to Se chalcogen vacancies that form during the growth process, many 3D TIs are naturally $n$-type, with the chemical potential in the bulk conduction band. However, many utilizations of the surface states require having the chemical potential inside the band gap to limit the signal from the bulk, and this is usually achieved with chemical substitution.

Much of the previous work on bulk-insulating TIs centers on the \ce{Bi_{2-x}Sb_xTe_{3}} (BST) or \ce{Bi_{2-x}Sb_xTe_{3-y}Se_y} (BSTS) family of materials \cite{Peng2012,Xia2013,Onishi2015} based on \ce{Bi2Te3}, but comparatively fewer studies exist on bulk-insulating TIs based on \ce{Bi2Se3}, such as \ce{Bi_{2-x}Sb_xSe3} \cite{Plechek2002}. \ce{Bi2Se3} has a larger band gap, more isotropic surface states with a more ideal spin-texture, and a Dirac point that is well separated in energy from the bulk valence band \cite{Xia2009,Chen2009,Hsieh2009,Souma2011,Ando2013}. These materials' differences persist in insulating alloys \cite{Ko2013}, and can lead to phenomena manifesting uniquely or more robustly in \ce{Bi2Se3}-based materials \cite{McIver2011,Hou2019}. They can also be expected to produce in different bulk dynamics of optically excited carriers, which heretofore have not been characterized in bulk insulating \ce{Bi2Se3}-based TIs.

In this work we report dramatically enhanced photoexcitation lifetimes in \ce{Bi_{2-x}Sb_xSe3} nanoplatelets. This increased lifetime only exists in bulk-insulating samples and exhibits fluence, temperature, and Sb-doping dependence distinct from the behavior in bulk-metallic ($n$-type) \ce{Bi2Se3}. The fluence-dependent decay rate is consistent with the bimolecular recombination of electron-hole pairs in the TI bulk, and connections to the recently reported evidence of exciton condensation are discussed.

Time-resolved transient reflectivity is a pump-probe technique that is extremely agile in terms of the materials and phenomena it can access. A pump pulse creates excitations into unoccupied states, and a second probe pulse measures the transient change in reflectivity $\Delta R/R$ at a time delay $t_\text{delay}$ later. The magnitude of $\Delta R/R$ can be used as a proxy for the number of nonequilibrium excitations \cite{Gedik2004}, and its evolution $\Delta R(t_\text{delay})/R$ can reveal the processes by which these photoexcited carriers return to equilibrium. This technique has been successful in studying low-energy excitations (i.e., subexcitation frequency) in superconductors \cite{Averitt2001,Gedik2004,Torchinsky2010,Torchinsky2011}, charge density wave systems \cite{Yusupov2008}, correlated electron systems \cite{Chen2016}, and topological quantum materials \cite{Kumar2011,Cheng2014,Glinka2015}.

Pulses are generated using a mode-locked, Ti:Sapphire oscillator (\SI{80}{\femto\second} pulse duration, \SI{80}{\mega\hertz} repetition rate, $E_\text{pump}=$ \SI{1.55}{\electronvolt}). The deposited energy per unit area, the pump fluence $\Phi$, is selected with neutral density filters, and $t_\text{delay}$ is controlled by a retroreflector mounted on a mechanical delay line in the optical path of the probe. The pump and probe pulses are focused to a single spot with a diameter of $d=$ \SI{40}{\micro\meter} at the sample surface. The probe reflection $R$ is measured by a photodiode, and the pump-induced $\Delta R$ signal is obtained with standard lock-in detection. The pump and probe pulses are cross-polarized to minimize interference at the sample surface and to limit pump scatter incident on the photodiode. Due to the penetration depth of \SI{1.55}{\electronvolt} light into \ce{Bi2Se3} ($\alpha\sim$ \SI{24}{\nano\meter}) \cite{McIver2012}, the experiment primarily samples the bulk. Further evidence that the bulk dominates our signal comes from time-resolved angle-resolved photoemission spectroscopy (trARPES) measurements showing softening of a characteristic coherent phonon at the surface, whereas only the original frequency is seen optically \cite{Sobota2014_PRL}.

We study \ce{Bi_{2-x}Sb_xSe3} nanoplatelets grown by chemical vapor deposition (CVD) and \ce{Bi2Se3} nanoplatelets synthesized from the same precursors. Typical nanoplatelet dimensions are $150 \times 150 \times \SI{0.1}{\micro\meter\cubed}$. The thickness of the specimens is well outside the regime where hybridization between opposite surfaces leads to the opening of a gap at the Dirac point \cite{Zhang2010, Weis2017}. Sb composition ranges from $x = 0.22 \text{to} 0.34$ for the \ce{Bi_{2-x}Sb_xSe3} samples, determined with energy-dispersive x-ray spectroscopy (EDS), which also correlates with carrier density \cite{Devidas2017}. After synthesis, the nanoplatelets are transferred by Kapton-tape onto a Si substrate capped with \SI{300}{\nano\meter}-thick \ce{SiO2}. No signal from the bare substrate is observed for the fluences used in this work.

\begin{figure}[h]
\includegraphics[width=0.5\textwidth]{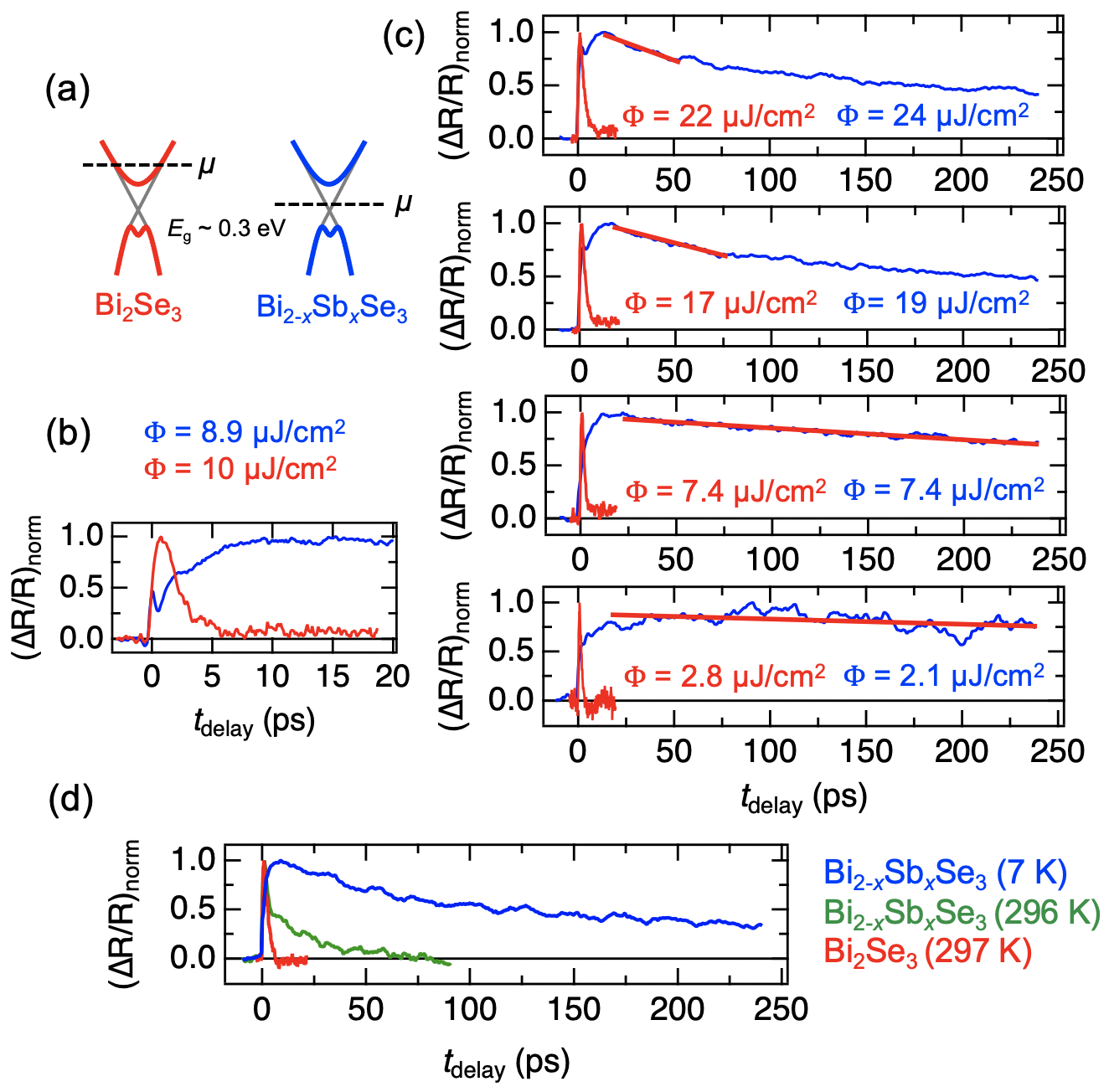}
\caption{Different relaxation dynamics in \ce{Bi2Se3} (red) and \ce{Bi_{2-x}Sb_xSe3} (blue). (a) Band diagrams of \ce{Bi2Se3} and \ce{Bi_{2-x}Sb_xSe3}, with $\mu$ denoting the schematic position of the chemical potential. (b) Example short delay time $\Delta R/R$ traces, normalized to their respective peak values at $T = \SI{7}{\kelvin}$. (c) Longer delay time traces with the same samples as in panel (b) at selected pump fluences at $T = \SI{7}{\kelvin}$. The red lines overlaying the blue traces show linear fits to the decays for $(\Delta R/R)_\text{norm}>0.7$. (d) $(\Delta R/R)_\text{norm}$ traces for a different \ce{Bi_{2-x}Sb_xSe3} sample at \SI{7}{\kelvin} (blue) and \SI{296}{\kelvin} (green), with \ce{Bi2Se3} at \SI{297}{\kelvin} (red) shown for comparison.}
\label{fig_1}
\end{figure}

Figure \ref{fig_1} shows the normalized transient reflectivity traces for bulk-metallic \ce{Bi2Se3} (red) and bulk-insulating \ce{Bi_{2-x}Sb_xSe3} nanoplatelets (blue, $x=0.25$) at \SI{7}{\kelvin} for several different fluences.  $t_\text{delay}=0$ corresponds to the time that the pump and probe pulses are coincident. Upon reaching their peak, the \ce{Bi2Se3} transient reflectivity traces decay to equilibrium in $\tau\sim $ \SI{2}{\pico\second}, consistent with prior transient reflectivity work on bulk \ce{Bi2Se3} \cite{Qi2010,Kumar2011,Lai2014,Glinka2015,Jnawali2018}. In contrast, the \ce{Bi_{2-x}Sb_xSe3} traces in Fig. \ref{fig_1}\textcolor{blue}{(c)} show much longer-lived excitations with $90\%$ of the reflectivity surviving near the edge of the measurement window at the lowest fluence $(\Phi=$ \SI{2.1}{\micro\joule\per\centi\meter\squared}). The transient reflectivity of the \ce{Bi_{2-x}Sb_xSe3} samples has a pronounced fluence dependence, which can be seen in the data in Fig. \ref{fig_1} by observing the magnitude of $(\Delta R/R)_\text{norm}$ near \SI{250}{\pico\second}. Relaxation becomes faster at higher temperature, but even at room temperature, \ce{Bi_{2-x}Sb_xSe3} shows longer-lived excitations than \ce{Bi2Se3} (Fig. \ref{fig_1}\textcolor{blue}{(d)}, green). More details on the $\Delta R/R$ signal structure are given in the Supplemental Material (SM) \cite{SM}.


\begin{figure}[h]
\includegraphics[width=0.5\textwidth]{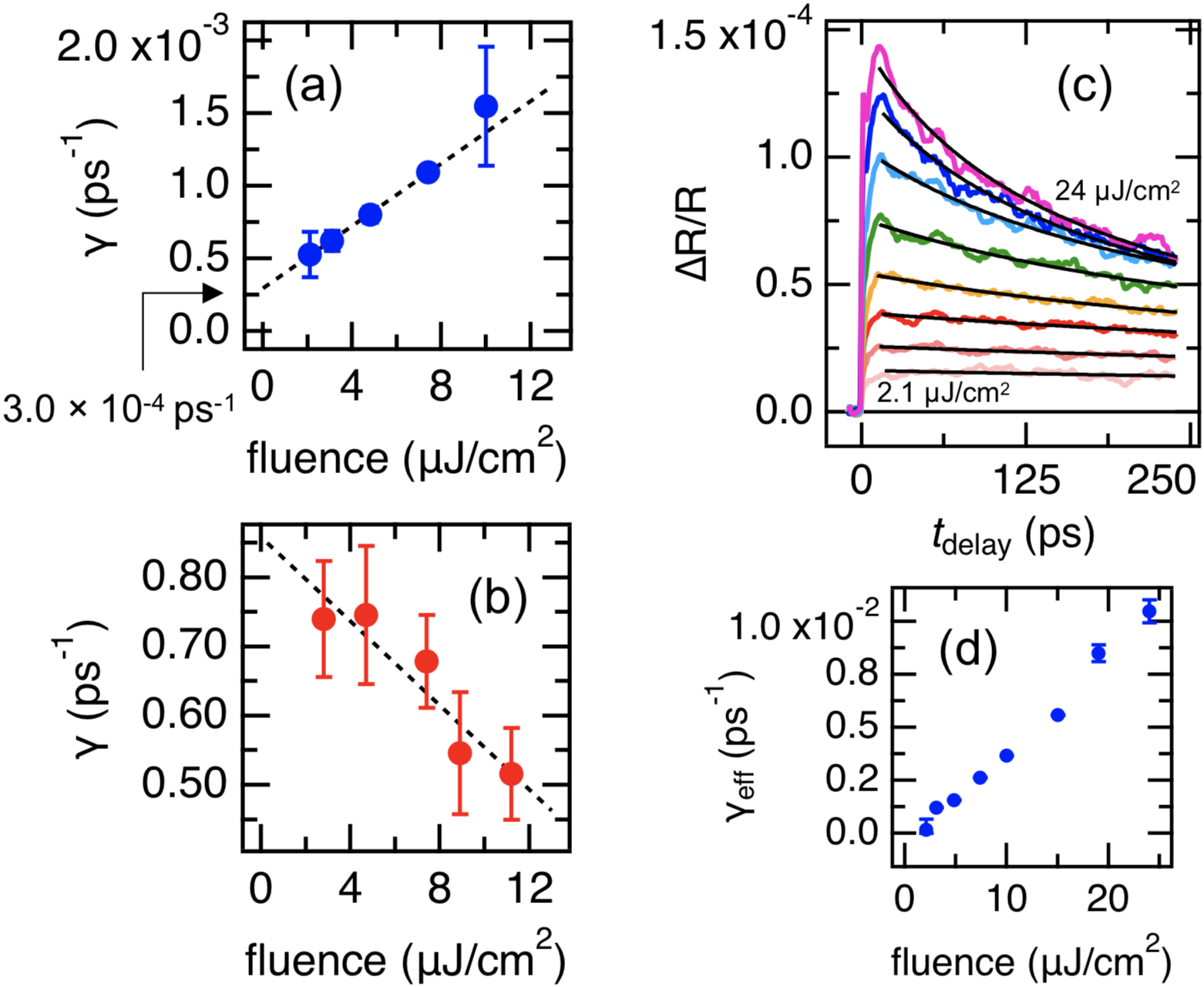}
\caption{\label{fig_2} Fluence dependence at $T = \SI{7}{\kelvin}$ in \ce{Bi_{2-x}Sb_xSe3} and \ce{Bi2Se3}. (a) Decay rates $\gamma$ for \ce{Bi_{2-x}Sb_xSe3} from linear fits of $(\Delta R/R)_\text{norm}$ vs $t_\text{delay}$. Dashed line: Linear fit to $\gamma$ vs fluence. (b) Decay rates for \ce{Bi2Se3} from single exponential fits as a function of fluence. (c) \ce{Bi_{2-x}Sb_xSe3} traces at different fluences, overlaid by fits to Eq. \ref{eqn1} (black). (d) Extracted bimolecular decay rates $\gamma_{\text{eff}}$ from the fits to Eq. \ref{eqn1}.}
\end{figure}

Figure \ref{fig_2} quantifies the fluence dependences for \ce{Bi_{2-x}Sb_xSe3} and \ce{Bi2Se3} at \SI{7}{\kelvin}. For \ce{Bi_{2-x}Sb_xSe3} at lower fluences, we quantify the decay rate using linear fits to the normalized traces over intervals where the decay is linear in time (red lines in Fig. \ref{fig_1}\textcolor{blue}{(c)}). The decay rates $\gamma$ from these fits are plotted in Fig. \ref{fig_2}\textcolor{blue}{(a)} as a function of fluence for \ce{Bi_{2-x}Sb_xSe3}. This model-independent fitting shows that decay rates become slower as the number of excitations (as parametrized by the fluence) is reduced, and suggests a linear relationship between the two. This linear relationship is characteristic of a bimolecular recombination process. Extrapolating to the zero fluence limit, the \ce{Bi_{2-x}Sb_xSe3} samples yield a decay rate of $\gamma (\Phi\rightarrow0)\sim$ \SI{0.30}{\per\nano\second}, or a decay time of \SI{3.3}{\nano\second}, substantially slower than in metallic \ce{Bi2Se3}. In contrast, the decay rates, derived from single-exponential fits, for \ce{Bi2Se3} in Fig. \ref{fig_2}\textcolor{blue}{(b)} show a fluence dependence \textit{opposite} of that of \ce{Bi_{2-x}Sb_xSe3}. Data on $n$-type \ce{Bi2Se3} samples show a plateau at sufficiently long delay times ($>\SI{10}{\pico\second}$), which increases with fluence, indicating steady state heating \cite{Kumar2011}. This plateau reaches $<10 \%$ of the maximum $\Delta R/R$ for $\Phi < \SI{22}{\micro\joule\per\centi\meter\squared}$, indicating minimal steady state heating in the fluence regime in Fig. \ref{fig_2} where our analysis of insulating samples is performed.

Following these model-independent observations, we now turn to a fitting scheme which specifically assumes bimolecular recombination (i.e., electron-hole recombination across the bulk band gap) to describe the decay dynamics in \ce{Bi_{2-x}Sb_xSe3} \cite{Gedik2004,Torchinsky2010,Torchinsky2011}. This fitting also incorporates the exponential decay of the pump and probe pulses in the sample, which implies the generation of a nonuniform $\propto n_0e^{-\alpha z}$ excitation density. This is accounted for with the following function for the transient reflectivity \cite{Gedik2004}:
\begin{equation}\label{eqn1}
\frac{\Delta R(t)}{R}=\frac{\Delta R(0)}{\gamma_{\text{eff}} t} \left[1-\frac{\ln(1+\gamma_{\text{eff}} t)}{\gamma_{\text{eff}} t}\right].
\end{equation}

The fitting parameters are $\Delta R(0)$ and $\gamma_{\text{eff}}$, the initial reflectivity change and the effective decay rate for the bimolecular process. Here, the decay rate is defined as $|\gamma_{\text{eff}}|\equiv\beta n$, with a quasiparticle density $n$ and a coefficient for the bimolecular process $\beta$. By accounting for the depth-dependent fluence, we can extend the applicability of the bimolecular recombination model to higher fluences. These traces and their fits to Eq. \ref{eqn1} are plotted in Fig. \ref{fig_2}\textcolor{blue}{(c)}, and the extracted fluence-dependent rates $\gamma_{\text{eff}}$ are then plotted in Fig. \ref{fig_2}\textcolor{blue}{(d)}.

\begin{figure}[h]
\includegraphics[width=0.5\textwidth]{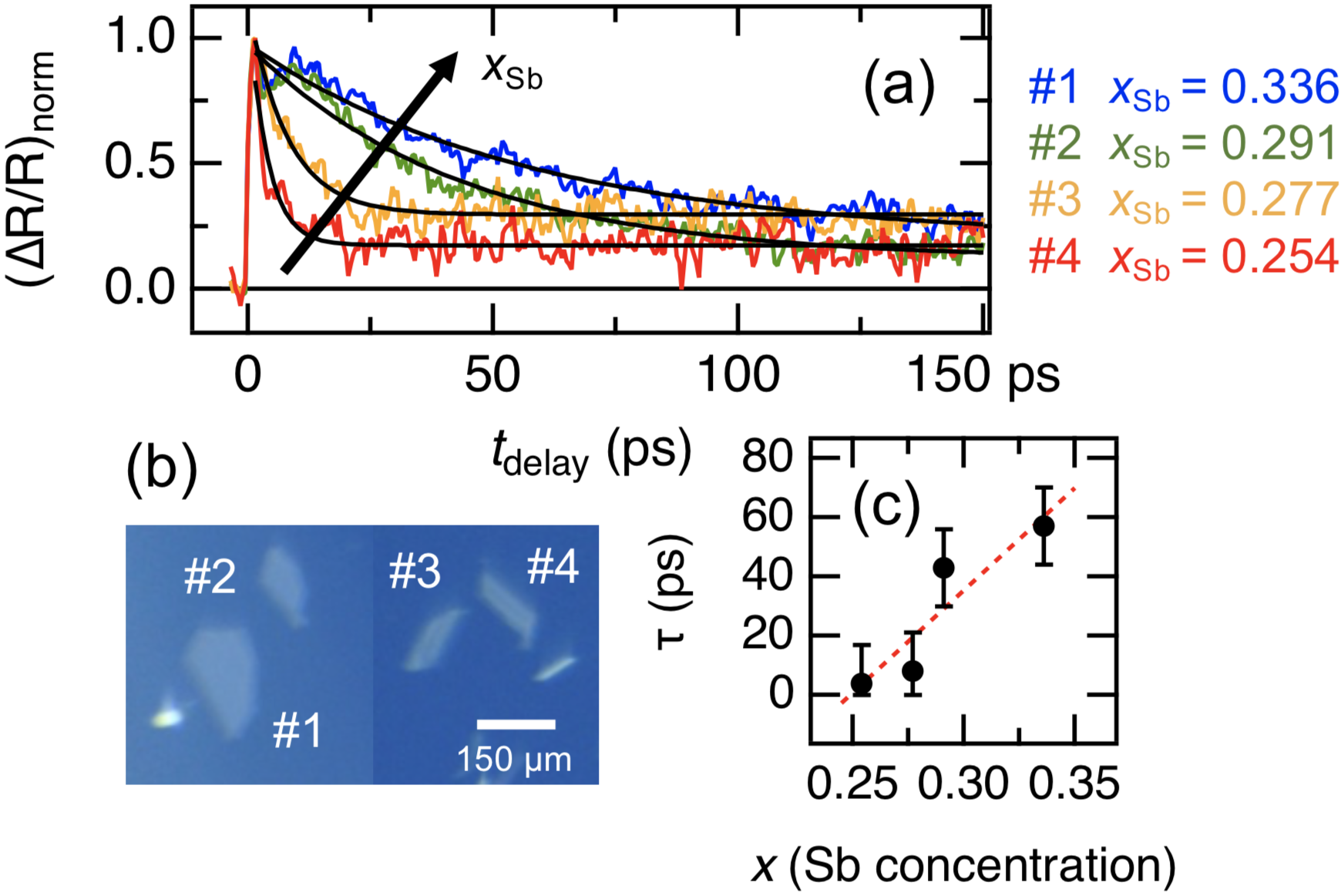}
\caption{Doping dependence of relaxation dynamics in \ce{Bi_{2-x}Sb_xSe3}. (a) $(\Delta R/R)_\text{norm}$ traces at $T = \SI{7}{\kelvin}$ and $\Phi = \SI{10}{\micro\joule\per\centi\meter\squared}$, with increasing Sb concentration denoted by the arrow. (b) Corresponding sample images. (c) Extracted time constants from $(\Delta R(t_\text{delay})/R)_\text{norm} = Ae^{-t_\text{delay}/\tau}+B$ fits with the dashed line indicating the overall trend.}
\label{fig_3}
\end{figure}

Figure \ref{fig_3}\textcolor{blue}{(a)} shows the variation in the $\Delta R/R$ traces as a function of Sb-doping, taken at a constant pump fluence. The specimens in this figure are different nanoplatelets from a single growth. To quantify decay times, the traces are fitted to single exponentials $(\Delta R(t_\text{delay})/R)_\text{norm} = A e^{-t_\text{delay}/\tau} + B$. The time constants from the fits are plotted in Fig. \ref{fig_3}\textcolor{blue}{(c)} as a function of $x$, the Sb concentration in \ce{Bi_{2-x}Sb_xSe3}. As $x$ increases, the carrier lifetime monotonically increases.

We begin by discussing bulk-metallic \ce{Bi2Se3}. The initial decay of $\Delta R/R$ in \ce{Bi2Se3} follows the framework of the two-temperature model for metals, where an out-of-equilibrium population of carriers with temperature $T_e$ thermalizes with the lattice with temperature $T_l$ by coupling to multiple phonon modes \cite{Allen1987,Groeneveld1995,Qi2010,Wang2012,Sobota2014_PRL,Lai2014}. The fast initial bulk carrier decay in \ce{Bi2Se3} is consistent with previous transient reflectivity measurements \cite{Qi2010,Kumar2011,Lai2014,Glinka2015,Jnawali2018} as well with transient \SI{}{\tera\hertz} conductivity work in thin films \cite{Sim2014,ValdsAguilar2015}. Thus, we can generalize the response of our \ce{Bi2Se3} samples as that of a typical metal. As noted earlier, the rates $\gamma$ for \ce{Bi2Se3} in Fig. \ref{fig_2}\textcolor{blue}{(b)} follow an opposite fluence dependence from the insulating samples, which underlines the importance of tuning the chemical potential for influencing the TI's response to optical excitation.

We now turn to bulk-insulating \ce{Bi_{2-x}Sb_xSe3}. As with metallic \ce{Bi2Se3}, photoexcitation at $t_\text{delay}$= \SI{0}{\pico\second} causes electrons from within the bulk valence band to populate bulk states far above $E_F$, but unlike the $n$-type system, electrons relax to the edge of the bulk conduction band which is minimally occupied at low temperature. This rapid initial 1-5 \SI{}{\pico\second} thermalization has been verified by trARPES \cite{Crepaldi2013,Sobota2014_ARPES,Sobota2014_PRL,Sterzi2017,Sumida2017,Freyse2018}, and thus the optical pump has the effect of an indirect injection of gap-energy excitations. The observed linear fluence dependence of the decay rate in  \ce{Bi_{2-x}Sb_xSe3} is consistent with bimolecular recombination, where photoexcited electron-hole pairs of density $n$ recombine and follow the simple rate equation: $\frac{dn}{dt}=-\beta n^2$, which when integrated yields a decaying quasiparticle density $n(t)=\frac{n_0}{1+n_0 \beta t}$ \cite{Gedik2004,Torchinsky2010,Torchinsky2011}. The effective decay rate $\gamma_{\text{eff}}$ then takes the form $\gamma\equiv\frac{1}{n}\frac{dn}{dt}=-\beta n$, where $\beta$ is the bimolecular coefficient. This recombination is assumed to be radiative because the band gap of \ce{Bi2Se3} ($E_g \sim \SI{0.3}{\electronvolt}$) is much higher than the highest phonon energy ($E = \SI{23}{\milli\electronvolt}$) in the material \cite{Zhang2011,Glinka2015}.

For greater context, a summary of time-resolved optical, mid-IR, and THz studies that have reported bulk photoexcited carrier lifetimes in \ce{Bi2Se3}-related compounds \cite{Qi2010,Kumar2011,Luo2013,Sim2014,Cheng2014,Lai2014,Onishi2015,ValdsAguilar2015,Glinka2015,Jnawali2018} is shown in the SM \cite{SM}. Neither long bulk carrier lifetimes in excess of \SI{1}{\nano\second} nor a fluence-dependence characteristic of bimolecular recombination has been reported simultaneously in those prior studies.  Thus, our measured long bulk carrier lifetimes in insulating samples, combined with the observed strongly fluence-dependent carrier recombination, points to a distinct interpretation of bulk recombination dynamics in insulating 3D TIs.

TrARPES studies have reported long-lived carriers arising from bulk excitations relaxing through the metallic surface states or from surface photovoltage (SPV) in 3D TIs \cite{Hajlaoui2014,Sobota2014_ARPES,Sterzi2017,Sumida2017,Freyse2018,Ciocys2019_stat_mech,Ciocys2020}. While the former may be the dominant relaxation mechanism near the surface, it cannot produce the strong fluence dependence we observe for \ce{Bi_{2-x}Sb_xSe3}. The surface state has a limited density of states near the Dirac point which restricts faster relaxation at higher excitation-densities. We note that the decay of the SPV has the opposite fluence dependence from our results in Fig. \ref{fig_2}\textcolor{blue}{(a)} \cite{Ciocys2020}.

\begin{figure}[t]
\includegraphics[width=0.5\textwidth]{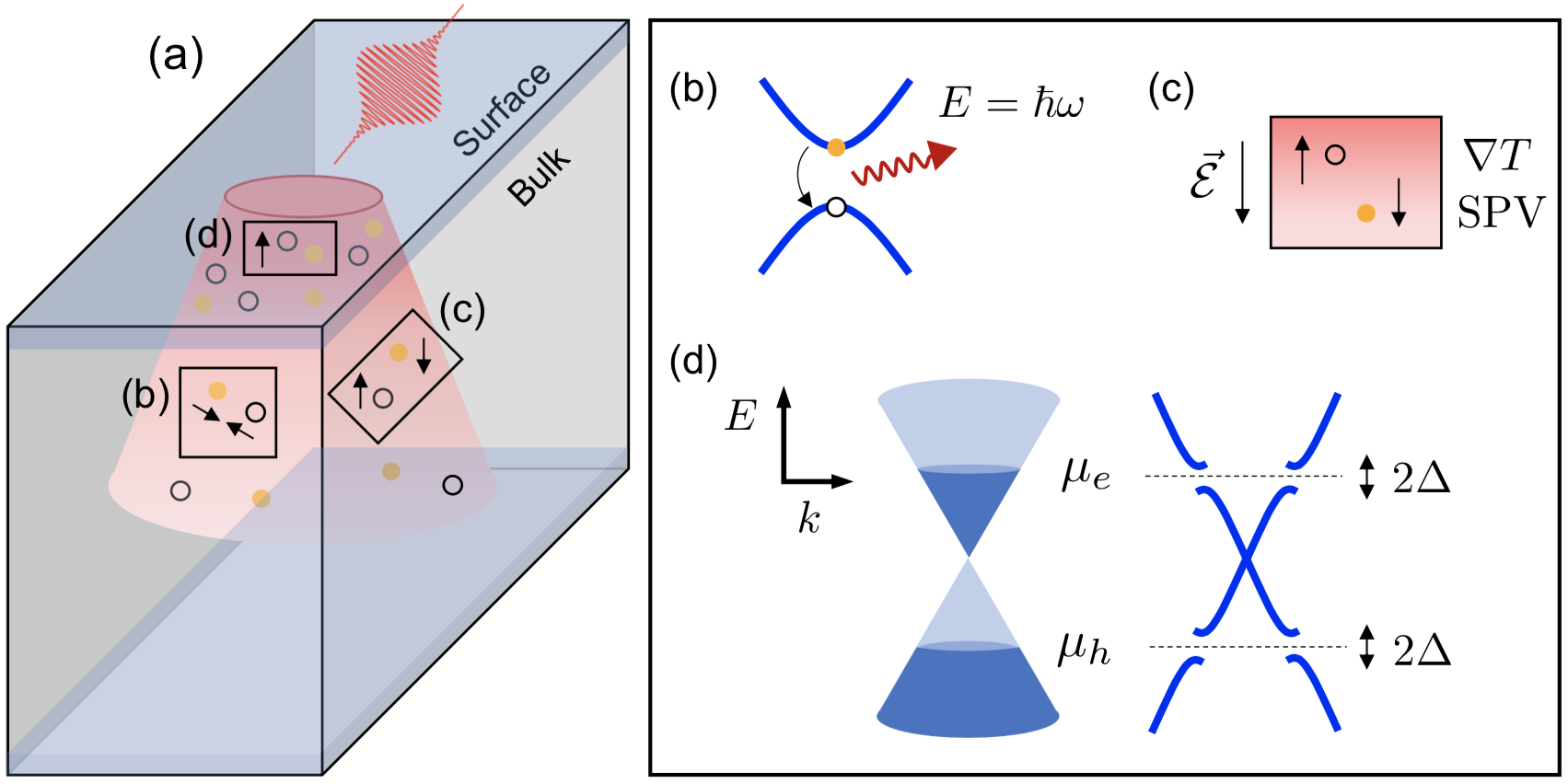}
\caption{\label{fig_4} Schematic of decay processes for \ce{Bi_{2-x}Sb_xSe3}. (a) Optical excitation generates electron-hole pairs in the bulk and surface of the TI. (b) Radiative, bimolecular recombination of electrons and holes. (c) Carrier migration due to a temperature gradient $\nabla T$ and surface photovoltage (SPV). (d) Long-lived carriers in surface Dirac state form different chemical potentials for electrons and holes ($\mu_e, \mu_h$) (left) where excitonic gaps ($2\Delta$) can open (right).}
\end{figure}

Additionally, the bulk decay dynamics of \ce{Bi_{2-x}Sb_xSe3} are highly doping-dependent, as illustrated by the traces in Fig. \ref{fig_3}, which may be interpreted either in terms of a changing free carrier density or doping inhomogeneity. Sb-doping on the Bi site is isovalent, and thus does not introduce charge carriers directly. Instead, it results in a smaller unit cell which is thought to diminish the number of Se vacancies in the studied doping range \cite{Plechek2002,Devidas2017}. In this doping regime, the free carrier density decreases monotonically with increasing Sb concentration $x$ \cite{Plechek2002}. These carriers are primarily thermally excited electrons in the conduction band, and thus, lower values of $x$ correspond to more recombination opportunities for photoexcited holes.  Thus, the observed doping dependence is consistent with a bimolecular recombination model.

Another relevant aspect is a possible distribution of local dopings, primarily due to varying Se vacancy density, which has been described in similar
nanoplatelets \cite{Lu2017,Lewin2018}. In samples closer to the $n$-type regime (smaller $x$), small local deviations from the average doping are more likely to correspond to local metallicity. As shown earlier, metallic and insulating \ce{Bi2Se3} yield profoundly different decay dynamics, and an increased probability of locally higher carrier density may promote behavior more like the former. We note some variation between different growths, which presumably arises from precursor variability in the CVD synthesis process, and the samples in Fig. \ref{fig_1} were from a different growth.

A summary of relaxation and migration processes in the bulk and surface regions in \ce{Bi_{2-x}Sb_xSe3} is shown in Fig. \ref{fig_4}, combining our results with those from literature. The pump initially generates electron-hole pairs in the bulk and surface regions of the material, which can subsequently undergo several processes during the decay to equilibrium. The first process for bulk carriers is bimolecular recombination across the bulk band gap. Near the band edge, bulk carriers recombine and release energy radiatively. The long nanosecond lifetime of this process allows time for other processes, such as photothermoelectric effects and SPV to assist the remaining carriers in migrating towards the surface \cite{Schroder2001,Yoshikawa2018,Ciocys2019_stat_mech,Hou2019}. In both cases, either a temperature gradient \cite{Sterzi2017} or a photodoping gradient produces an internal voltage that can sweep carriers from the deep bulk towards the surface (Fig. \ref{fig_4}\textcolor{blue}{(c)}).  When carriers ultimately relax, they slowly recombine and return to equilibrium after $t_\text{delay}>$ \SI{3}{\nano\second}, well outside our measurement window.

Near the surface, trARPES measurements show that bulk carriers at the surface slowly relax through the surface state (Fig. \ref{fig_4}\textcolor{blue}{(d)}, left) often establishing a long-lived carrier population at the surface \cite{Hajlaoui2014,Sobota2014_ARPES,Sterzi2017,Sumida2017,Freyse2018}. Importantly, these long-lived surface excitations are only observed in bulk-insulating or $p$-type samples (e.g., lifetimes exceeding \SI{400}{\pico\second} in \ce{(Sb_{1-x}Bi_x)2Te3} \cite{Sumida2017} and in Mg-doped \ce{Bi2Se3} \cite{Ciocys2020}). Our transient reflectivity experiments contribute nuance to this picture by illustrating the behavior of bulk carriers away from the surface, which are excited at the same time and can populate the surface state at later times because they are even more long-lived.

The decay bottleneck near the surface allows electrons and holes to develop separate chemical potentials $\mu_e,\mu_h$ (Fig. \ref{fig_4}\textcolor{blue}{(d)}, left), relevant to an important prediction in TIs: exciton condensation \cite{Seradjeh2009,Cho2011,Tilahun2011,Moon2012,Mink2012,Efimkin2012,Rist2013,Triola2017,Hou2019}. Dirac materials, including 3D TIs, are predicted to allow the formation of an exciton condensate—a Bardeen-Cooper-Schrieffer-like ground state of bound electron-hole pairs at the TI surface with gating or optical excitation \cite{Seradjeh2009,Triola2017}. Key spectroscopic signatures of this state are excitonic gaps $2\Delta$ that form in the surface states at the chemical potentials $\mu_e,\mu_h$ (Fig. \ref{fig_4}\textcolor{blue}{(d)}, right) \cite{Triola2017,Pertsova2020}. These gaps have an estimated magnitude up to $\sim 1-3$ $\SI{}{\milli\electronvolt}$. Bimolecular recombination of carriers across these excitonic gaps would also produce the presently observed fluence dependence, and ultrafast optics is typically sensitive to recombination across \SI{}{\milli\electronvolt}-magnitude gaps with an \SI{}{\electronvolt}-magnitude probe \cite{Demsar1999,Averitt2001,Gedik2004,Yusupov2008,Torchinsky2010,Torchinsky2011,Chen2016}. However, as discussed earlier, it is likely that our signal is dominated by the bulk.

A recent study on the same samples as in the present work reported highly nonlocal, millimeter-long surface photocurrents in bulk-insulating \ce{Bi_{2-x}Sb_xSe3} after optical excitation \cite{Hou2019}, which have been interpreted in terms of exciton condensation. Tuning the chemical potential into the bulk gap is necessary for observing long relaxation times as it is for observing long decay lengths \cite{Hou2019,Hou2020}. Photocurrent decay lengths in \ce{Bi_{2-x}Sb_xSe3} are maximized at low temperature ($T<$ \SI{40}{\kelvin}) and low fluence, precisely the regime where we observe the longest relaxation times. Importantly, our long $\tau>\SI{3}{\nano\second}$ lifetime, observed in the fluence regime relevant to potential exciton condensation, is incompatible with millimeter-long diffusive carrier travel, which would imply a carrier mobility of $\mu> 10^5$ $\text{m}^2/\text{V}\cdot\text{s}$, much higher than the highest measured values of $\mu\sim 1$ $\text{m}^2/\text{V}\cdot\text{s}$ \cite{Butch2010,Xiu2011}.

In summary, our transient reflectivity results show a three orders-of-magnitude slowing of the carrier decay rate $\gamma\rightarrow\SI{0.30}{\per\nano\second}$ in the zero pump-fluence limit in photoexcited \ce{Bi_{2-x}Sb_xSe3} at $T=$ \SI{7}{\kelvin}, as compared to $n$-type specimens. Our fluence-dependent data reveal a distinct process to consider in bulk-insulating TIs: bimolecular recombination of bulk carriers after optical excitation. We additionally show the key role composition plays in both establishing long bulk carrier decay in bulk-insulating samples and in influencing carrier recombination rates. These findings underscore the role optically excited bulk carriers play in TIs prior to migrating to the surface, relevant for interpreting optoelectronic phenomena of surface states.

\begin{acknowledgments}
We thank Fahad Mahmood, Shuolong Yang, and Denis Golež for helpful discussions. This work is supported by National Science Foundation Grant No. DMR-1838532..
\end{acknowledgments}


\providecommand{\noopsort}[1]{}\providecommand{\singleletter}[1]{#1}%

\end{document}


\title{Nanosecond dynamics in intrinsic topological insulator \ce{Bi_{2-x}Sb_xSe3} revealed by time-resolved optical reflectivity: Supplementary material}

\maketitle
\section{Short and long time $\Delta R/R$ signal structure}

There is some complicated structure in the $\Delta R/R$ traces at short delay times, namely the delayed onset in the \ce{Bi_{2-x}Sb_xSe3} trace around $t_\text{delay}\sim$ \SI{0.5}{\pico\second} and the kink preceding the peak of the \ce{Bi2Se3} trace in Fig. S1 (left). These features have been observed in earlier transient reflectivity studies on \ce{Bi2Se3} \cite{Qi2010,Cheng2014} and attributed to a negative amplitude component arising from the trapping of electrons by Se vacancies \cite{Qi2010}. In our work, we can safely ignore the early structure because our primary focus is on the long-lived component to the decay for $t_\text{delay}\gg$ \SI{0.5}{\pico\second}, when any negative transients are expected to die out.

At longer time delays, there are also long-lived $f\sim\SI{40}{\giga\hertz}$ oscillations in the \ce{Bi_{2-x}Sb_xSe3} traces, which can most clearly been seen in Fig. S1 (right). These oscillations are attributable to acoustic phonon oscillations \cite{Qi2010,Kumar2011,Glinka2015}, and can only be resolved at higher pump fluences $(\Phi>\SI{10}{\micro\joule\per\centi\meter\squared})$.

\begin{figure*}[h]
\includegraphics[width=1\columnwidth]{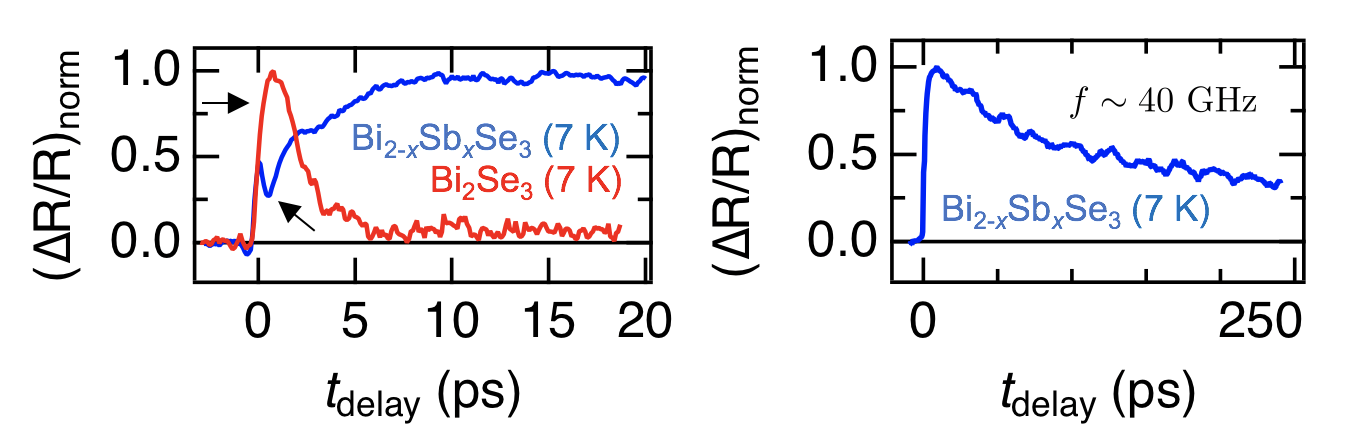}
\captionsetup{justification=centering,singlelinecheck=false}
	\caption{Left: Short delay time traces showing negative component and kink. Right: Long delay time trace showing acoustic phonon oscillations in \ce{Bi_{2-x}Sb_xSe3}.}
\end{figure*}

\section{Prior pump-probe works}
For greater context, there are several overlapping time-resolved optical, mid-IR, and THz works that have previously measured optically excited bulk carrier lifetimes in \ce{Bi2Se3}-based compounds. A selection of these works are listed in Table I, alongside the relevant experimental conditions, relaxation time, and the fluence-dependence of the carrier decay rate $\gamma$ (or scattering rate for the THz studies). This table focuses on optical, mid-IR, and THz measurements in order to compare bulk sensitive measurements to one another. ARPES studies are excluded from this table because they are more surface sensitive, but these works are cited and discussed in the main text. 

This list is not meant to be fully comprehensive, but rather to contextualize our findings within the broader landscape of works. Our measured long bulk carrier lifetimes in insulating samples, combined with the observed novel fluence dependence, distinguishes our study from these works.

\begin{figure*}[h]
\includegraphics[width=1\columnwidth]{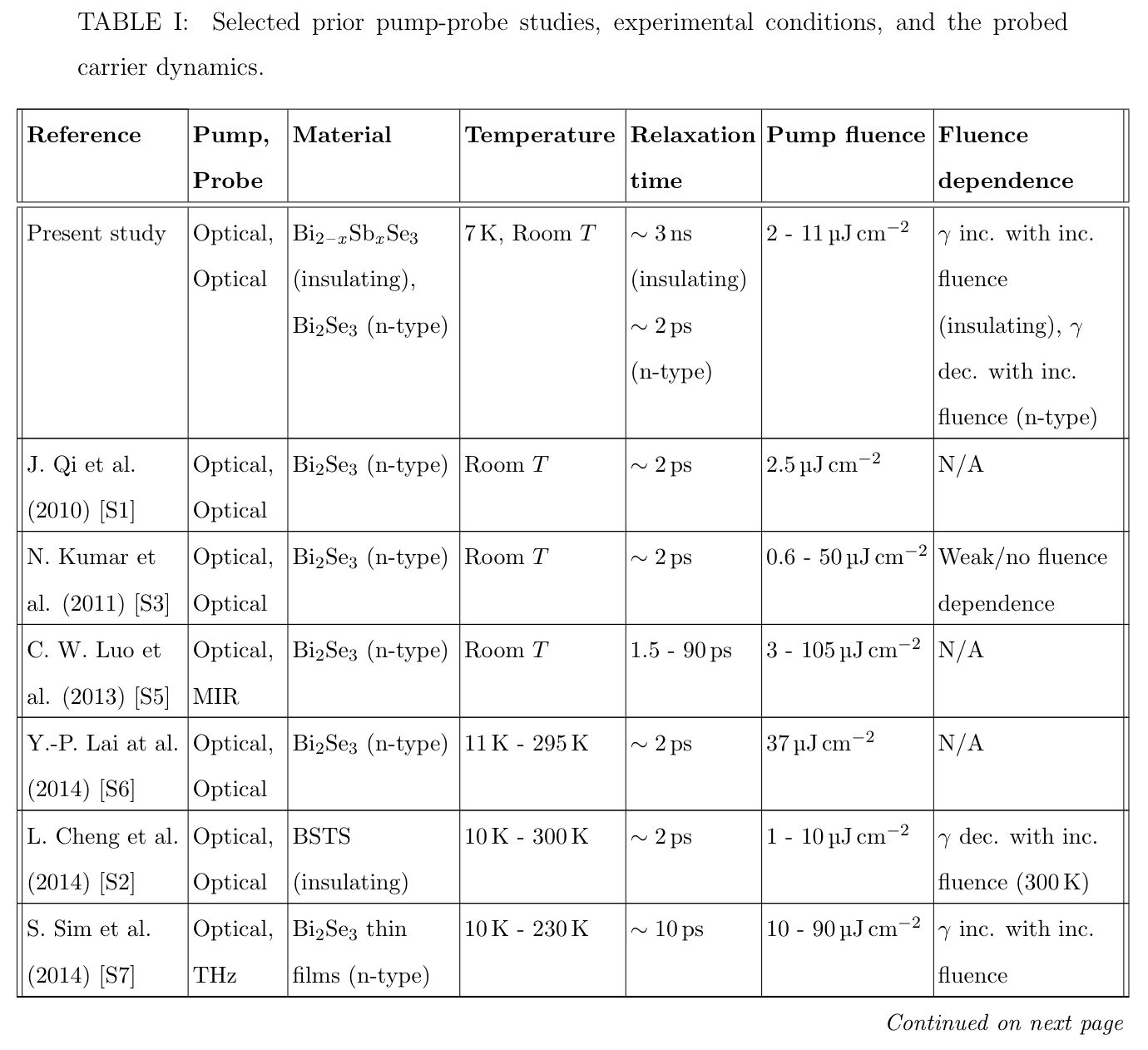}
\end{figure*}

\begin{figure*}[h]
\includegraphics[width=1\columnwidth]{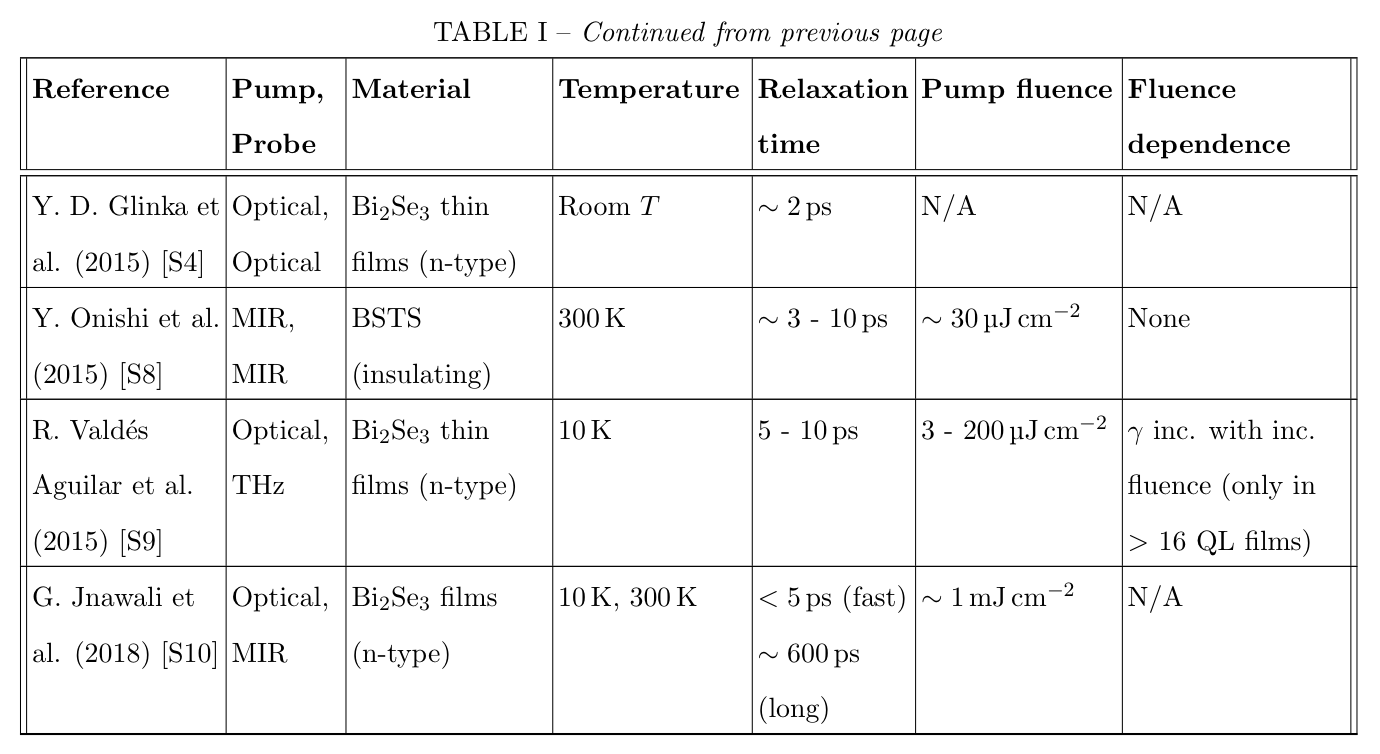}
\end{figure*}

\pagebreak

\providecommand{\noopsort}[1]{}\providecommand{\singleletter}[1]{#1}%